\documentclass{elsart1p}
\usepackage{graphicx}
\usepackage{amssymb}
\newcommand{\mrf}[1]{\mbox{$\mathrm{#1}$}}
\newcommand{\mif}[1]{\mbox{$\mathit{#1}$}}
\newcommand{\mat}[1]{\mbox{$#1$}}
\newcommand{\dtq}{\mat{\Delta_T q(x)}}
\newcommand{\dq}{\mat{\Delta q(x)}}

\newcommand{\qq}{\mat{q(x)}}

\newcommand{\gom}{\mrf{GeV/\mif{c}}}
\begin{document}
\begin{frontmatter}
%
%
%
\title{Transverse spin effects in COMPASS}
%
%
\author{A. Bressan, on behalf of the COMPASS Collaboration}
\address{Universit\`a degli Studi di Trieste and INFN - Sezione di Trieste,
  via Valerio, 2 - 34127 Trieste, Italy}
\begin{abstract}
In the years 2002-2004 COMPASS has collected data with the $^6$LiD target
polarization oriented transversely with respect to the muon beam direction for
about 20\% of the running time, to measure transverse spin effects in
semi-inclusive deep inelastic scattering, one of the main objectives of the
COMPASS spin program. In 2007, COMPASS has used for the first time a proton
NH$_3$ target with the data taking time equally shared between longitudinal and
transverse polarization of the target. After reviewing the results obtained with
the deuteron, the new results for the Collins and Sivers asymmetries of the
proton will be presented.
\end{abstract}
\end{frontmatter}
Since 2002 the COMPASS experiment has performed DIS measurements by impinging
160~\gom\ positive muons on a solid polarized deuteron or proton target, to get a
deeper insight into the structure of the nucleon. Till 2006 a \mrf{^6LiD}
target has been used, while a \mrf{NH_3} target was used in 2007. In 2008
COMPASS has started to measure central production and diffractive processes,
from the scattering of 190~\gom\ pions on a liquid \mrf{H_2} target, to search
for exotic hadronic states such as hybrids and glue-balls. 

The experimental apparatus consists of a two stage magnetic spectrometer able to
separate the scattered muon from the produced hadrons. Charged particles are
identified by a RICH and by hadronic calorimeters. Since 2006 a new target
magnet has been installed increasing the acceptance from 70 mrad scattering
angle up to 180 mrad. A complete description of the experiment can be found
in~\cite{nimcompass}. 
\section{Collins and Sivers asymmetries}
The full description of the spin structure of the nucleon at twist-two level
requires the knowledge of the transversity distributions \dtq, together with the
helicity distributions \dq\ and \qq. The transversity distributions \dtq\ are
difficult to measure since, being chirally odd, they decouple from inclusive deep
inelastic scattering (DIS).  However, they can be measured in semi-inclusive DIS
(SIDIS) in combination with chiral-odd fragmentation functions, such as the
Collins fragmentation function $\Delta^0_T D_q^h$ for hadron production, giving
rise to an azimuthal single spin asymmetry (SSA) in the final state
hadrons~\cite{Collins:1992kk}. The Collins fragmentation function describes the
spin-dependent part of the hadronization of a transversely polarized quark into
a hadron with transverse momentum $p^h_T$. At leading order, the Collins
mechanism leads to a modulation in the azimuthal distribution of the 
produced hadrons given by: 
\[
N(\Phi_{C}) = \alpha(\Phi_{C}) \cdot N_0 \, (1 + A\mrf{_{Col}} \cdot P_T
\cdot f \cdot D_{NN} \sin{\Phi_{C}})\, ,
\]
\noindent where $\alpha$ contains the apparatus efficiency and acceptance, $P_T$
is the target polarization, $D_{NN}$ is the spin transfer coefficient and $f$ is
the fraction of polarizable nuclei in the target; $\Phi_C =\phi_h-\phi_{S'} =
\phi_h+\phi_S - \pi$ is the Collins angle, with $\phi_h$ the hadron azimuthal
angle and $\phi_{S'}$ the final azimuthal angle of the quark spin and $\phi_S$
the azimuthal angle of the nucleon spin in the $\gamma-N$ system. Finally
\[
A\mrf{_{Col}} = \frac{\sum_q e^2_q \cdot \Delta_T q(x) \cdot  \Delta^0_T
  D_q^h(z,p_T^h)}{\sum_q e^2_q \cdot q(x) \cdot  D^h_q(z,p_T^h)}  
\]
\noindent is the Collins asymmetry, arising from the product of the transversity
distribution $\Delta_T q$ and the Collins fragmentation function $\Delta^0_T D_q^h$. 

A different mechanism~\cite{sivers} has also been suggested in the past as
possible source of single
spin asymmetries in the cross-section of unpolarized leptons impinging on transversely
polarized nucleons. Allowing for an intrinsic $\vec{k}_T$ dependence of the
quark distribution in the nucleon, a transverse spin polarization may induce a
left-right asymmetry in such a distribution, giving rise to a measurable `Sivers
asymmetry':
\[
A\mrf{_{Siv}} = \frac{\sum_q e^2_q \cdot \Delta^T_0 q(x,p_T^h) \cdot 
  D^h_q(z)}{\sum_q e^2_q \cdot q(x) \cdot  D^h_q(z)} 
\]
with a modulation expressed in terms of the Sivers angle $\Phi_S =
\phi_h-\phi_S$. Since in this case the unpolarized fragmentation functions are
known, the measurement of the Sivers asymmetry for both positive and negative
produced hadrons allows directly to extract the Sivers functions, if the measured
asymmetry are different from zero, while a zero result for an isoscalar target
like the $^6$LiD used in COMPASS can come both from a vanishing Sivers function or 
from a cancellation between $u$ and $d$ quark contributions. 
\section{Results}
The event selection requires standard DIS cuts, i.e. $Q^2 > 1\ \mrf(GeV)/c)^2$,
mass of the final hadronic state $W>5\ \mrf{GeV}/c^2$, $0.1 < y < 0.9$, and the
detection of at least one hadron in the final state.  For the detected hadrons
it is also required that:
\begin{itemize}
\item the fraction of the virtual photon energy carried is
$z=E_h/E_\gamma>0.2$ to select hadrons from the current fragmentation region;
\item
$p_T > 0.1\ \mrf{GeV}/c$ (where $p_T$ is the hadron transverse momentum with
respect to the virtual photon direction) for a better determination of the
azimuthal angle $\phi_h$. 
\end{itemize}

\begin{figure}[t]
\begin{center}
\includegraphics[width=0.7\textwidth]{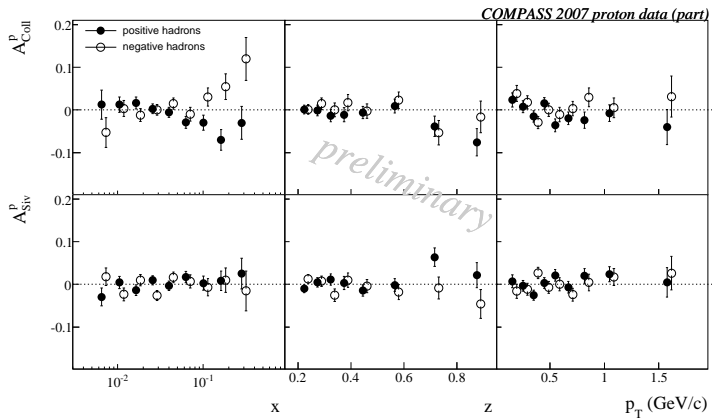}
\end{center}
\caption{Collins and Sivers asymmetry for positive (full points) and negative
  (open points) hadrons as a function of $x$, $z$ and $p_T$ for the proton 2007 data.}
\label{results}
\end{figure}

\noindent The asymmetries have been calculated as a function of $x$, $z$ and
$p_T$ for positive and negative hadrons respectively. 
Both the resulting Collins and Sivers asymmetries from the whole deuteron data
for all hadrons  turned out to be small and compatible with zero~\cite{npb}, a
trend that is also shown by the identified hadron results~\cite{plbid}, a
result which was interpreted as a cancellation between the contribution of the
$u$ and $d$ quarks, for the isoscalar deuteron target. The new results for the
proton \mrf{NH_3} target are shown in fig.~\ref{results} both for the Collins
(upper row) and for the Sivers (lower row) asymmetries. The Collins asymmetries
as a function of $x$ are small, compatible with 0, up to $x~\sim0.05$, while in
the last points a signal appears, and the asymmetries increases up to 10\% with
opposite sign for the positive (closed points) and negative (open points)
hadrons. The trend is in good agreement with what observed by
HERMES~\cite{hermes}. At variance the Sivers asymmetries are small and
compatible with zero over the full $x$ range and for both positive and negative
hadrons; in this case the compatibility with HERMES results is fine for negative
hadrons but is marginal, if any, for positive hadrons. The origin of the
disagreement, needs to be understood~\cite{vogelsang} and will be an interesting
issue for the near future.

\end{document}